\documentclass[aps, pra, preprint, superscriptaddress, amsmath, amssymb, showpacs]{revtex4-1}
\usepackage[english]{babel}
\usepackage[utf8]{inputenc}
\usepackage{longtable}
\usepackage{amssymb,bm}
\usepackage{graphicx}
\usepackage{amsmath}
\usepackage{epstopdf}
\usepackage{float}
\usepackage{amsfonts}
\usepackage{upgreek}
\usepackage{booktabs}

\begin{document}
	
	\title{Possibility to investigate P-parity violation in nuclear collisions at facility NICA}
	
	\author{I.A.\,Koop}
\email{{I.A.Koop@inp.nsk.su}}
\affiliation{Budker Institute of Nuclear Physics of SB RAS, 630090 Novosibirsk, Russia}
\affiliation{Novosibirsk State University, 630090 Novosibirsk, Russia}
\affiliation{Novosibirsk State Technical University,630092 Novosibirsk, Russia}

	\author{A.I.\,Milstein}
	\email{{A.I.Milstein@inp.nsk.su}}
	\affiliation{Budker Institute of Nuclear Physics of SB RAS, 630090 Novosibirsk, Russia}
	\affiliation{Novosibirsk State University, 630090 Novosibirsk, Russia}
	
	\author{N.N.\,Nikolaev}
\affiliation{L.D. Landau Institute for Theoretical Physics, 142432 Chernogolovka, Russia}

\author{A.S.\,Popov}
\affiliation{Budker Institute of Nuclear Physics of SB RAS, 630090 Novosibirsk, Russia}
\affiliation{Novosibirsk State University, 630090 Novosibirsk, Russia}

\author{S.G.\,Salnikov}
\affiliation{Budker Institute of Nuclear Physics of SB RAS, 630090 Novosibirsk, Russia}
\affiliation{Novosibirsk State University, 630090 Novosibirsk, Russia}	

\author{P.Yu.\,Shatunov}
\affiliation{Budker Institute of Nuclear Physics of SB RAS, 630090 Novosibirsk, Russia}
\affiliation{Novosibirsk State University, 630090 Novosibirsk, Russia}	

\author{Yu.M.\,Shatunov}
\affiliation{Budker Institute of Nuclear Physics of SB RAS, 630090 Novosibirsk, Russia}
\affiliation{Novosibirsk State University, 630090 Novosibirsk, Russia}

\date{\today}
	
	\begin{abstract}
		A possible experimental setup for measuring the effect of parity violation in the interaction of the polarized proton or deuteron beams with an unpolarized target is discussed. One possibility is investigation of  scattering of the proton or deuteron polarized beams on a thick internal target in one of the rings of the NICA collider. In this case, the spin of a circulating particles is transformed into a mode precessing in the horizontal plane using an RF flipper. The effect of parity violation will be studied by measuring the correlation of the interaction cross section of particles and the direction of their spins. In an alternative approach, the flipper transforms the spins of particles into a horizontal plane and the beam is extracted into the channel in a certain phase of the precession. In this more traditional experimental setup, the total cross section of the passage of particles through a dense target is measured, depending on the sign of the helicity of the polarization of the beam.
	\end{abstract}
	
	%\pacs{ 34.80.Nz}
	\maketitle

\section*{Introduction}

At present, the quantitative predictions of the Standard Model for electroweak processes obtained in the terms of quarks and leptons are in good agreement with experimental data for very large momentum transfers (at short distances). However, the analysis of rather extensive experimental data on parity violation in hadron interaction processes, which is associated with weak interactions, requires the use of phenomenology for small momentum transfers (at relatively large distances). Namely, it is necessary to use the  language of mesons and baryons (see ~ review \cite { Gardner2017xyl}). In the process of elastic or inelastic scattering of protons or deuterons on nuclei, the effect of parity violation will be manifested in the dependence of the total or differential elastic scattering cross section on the longitudinal polarization of the beam. This dependence arises due to interference of the amplitude corresponding to the strong interaction and the $ P $-odd amplitude corresponding to the weak interaction, which can be estimated quite reliably.

The most accurate proton-proton scattering experiment, devoting to the investigation of parity violation, was performed at 45\ MeV. The result for asymmetry,  $ A_L = - (1.5 \pm 0.22) {\cdot} 10 ^ { -7} $, was based on the data collected for ~ several years \cite{Kistryn1987tq}. The only experiment with a nonzero result for asymmetry,  $ A_L = - (26.5 \pm 6.0 \pm 3.6) {\cdot} 10 ^ {- 7} $,  was carried out for protons interacting with the water target at energies of 5.1 \, GeV  \cite {Lockyer1984pg}. This anomalously large asymmetry did not receive a satisfactory theoretical explanation. In the Fermi laboratory, a beam of polarized protons and antiprotons with an energy of 200 GeV, which was created as a result of decays of hyperons, had a low intensity and the achieved sensitivity to ~ $ P $-asymmetry was only at the level of $ 10^{- 5} $ \cite {Grosnick1996sy}.

In our work, we discuss several possibilities for experiments to observe the effect of parity violation in the scattering of polarized protons or light nuclei on an unpolarized nuclear target using the unique capabilities of the NICA  accelerator complex \cite{Kekelidze2016vcp, Savin2014sva} which is currently under construction  at ~ JINR.

\section{Method of precessing spin}

In this method, a polarized beam in the storage ring comes into interaction with a solid-state internal target using dynamic orbit distortion. In this case, events  are detected by scintillation counters located outside the vacuum chamber at a small distance from the target. The initial polarization is the equilibrium vertical polarization. Before the beam begins to interact with the target, the spin of the particles in the beam is adiabatically rotated from vertical to horizontal position with the help of the radio frequency flipper, then the flipper is turned off and free spin precession begins in the horizontal plane of the storage ring. The observed signal will be the oscillating component at ~ the count rate, proportional to the longitudinal component of the precessing polarization, the orientation of which is tracked by the time registration  of events. The rate of beam dumping onto a target is limited only by the counting speed of the detectors and the operating conditions of the target and may be fractions of a second. During this time, the spin manages to make hundreds of thousands of revolutions in the ~ horizontal plane, which will allow a qualitative Fourier analysis of the events. 

A similar scheme was used in JEDI experiments to study the coherence time of free spin precession of vector-polarized deuterons in the ~ COSY storage ring in ~ Julich \cite {Stephenson2015dkk}. In the  experiment  \cite{Stephenson2015dkk}  , deuterons were orbited to the internal carbon target of an EDDA polarimeter with a thickness of 17 mm. The signal in this experiment was up-down asymmetry in elastic scattering, proportional to the oscillating horizontal component of polarization perpendicular to the beam axis, and the effectiveness of the time-marking of events was proved. In a subsequent experiment \cite {Bagdasarian2014ega} it was shown that the coherence time of the horizontal spins of a deuteron beam can be brought to a thousand or more seconds. Note that acceleration of beams with precessing horizontal polarization was previously discussed in \cite {Sitnik2002}. The first study of the problem of increasing the coherence time of the free precession of an ensemble of horizontal spins was carried out in the experiment \cite {Vasserman1987cf}, in which the anomalous magnetic moments of the electron and positron were compared. It was shown \cite {Koop1988} that the coherence time of free precession can be significantly increased with the help of sextupole lenses. However, in the ~ proposed experiment for NICA, such a long coherence time is not required.

The azimuthally integrated elastic scattering cross section is independent of the transverse polarization of protons. However, with incomplete azimuthal symmetry of the detector, a residual dependence of the cross section on transverse polarization is possible. In our case, it will be a component in the counting speed, proportional to the oscillating component of the horizontal spin, perpendicular to the axis of collision. Since it will be difficult to achieve complete azimuthal symmetry of the detectors, we propose a method for suppressing this effect. Namely, the indicated background effect can be eliminated with the help of a Fourier analysis of the count rate, using the fact that the phase modulation of the effects associated with the longitudinal and transverse polarization is shifted by $ 90^\circ $. In addition, the continuous separation of azimuthal asymmetry is a way to control the magnitude of the horizontal polarization.
The signal in the upper counter $ U_1 $ and the lower counter $ U_2 $ can be conditionally represented in the form:
\begin{align*}
U_1 &= \left\langle U_1 \right\rangle \left[ 1 + A \cos(\psi + \mu n) + B \sin (\psi + \mu n) \right] \,,\\
U_2 &= \left\langle U_2 \right\rangle \left[ 1 - A \cos(\psi + \mu n) + B \sin (\psi + \mu n) \right].
\end{align*}
The terms proportional to $ A $ are related to the spin-orbit interaction and exist without weak interaction effects. The terms proportional to $ B $ are related to the weak interaction, $ \psi $ is the initial phase of spin precession in the ~ horizontal plane, $ \mu $ is the phase shift per revolution of the particle in the accelerator, $ n $ is the revolution number. The difference in the signals averaged over the precession phase, $ \left \langle U_1 \right \rangle $ and $ \left \langle U_2 \right \rangle $, is associated with a ~ possible asymmetry of registration efficiency. Therefore, the parameters $ A $, $ \psi $ and $ \mu $ can be extracted with ~ good accuracy from measuring
$$R = \dfrac{U_1}{\left\langle U_1 \right\rangle} - \dfrac{U_2}{\left\langle U_2 \right\rangle} = 2 A \cos(\psi + \mu n)$$
~ for many thousands of revolutions. After that, the coefficient $ B $, which is associated with the ~ parity violation effect, can be determined from the measurement of the sum
 $$S = \dfrac{U_1}{\left\langle U_1 \right\rangle} + \dfrac{U_2}{\left\langle U_2 \right\rangle} = 2 B \sin(\psi +\mu n).$$
The relative efficiency of the counters can be calibrated with the use of a ~ non-polarized beam, and by such a correction it is possible to significantly suppress the dependence of the total count rate on the polarization perpendicular to the collision axis. The degree of suppression can be estimated experimentally from the dependence of the counting rate on the sign of vertical polarization before and after correction for efficiency (in the experiment \cite {Brantjes2012zz}, a similar problem was solved to eliminate systematic effects for measuring the vector polarization of deuterons with an accuracy of $ 10^ {- 6 } $).
The signal from the longitudinal component of the spin is also contained in ~ two side counters. Using the same reference to the ~ phase of the spin precession, we can extract the signal from the longitudinal component of the spin from the sum of signals in these counters.

\section{Use of  flipper}
Let us discuss in more detail the experimental scheme based on the use of an ensemble of circulating in a ~ horizontal plane spins of the particles rotating in the storage ring. Our goal is to measure the effect of parity violation at the level of $ 10 ^ {- 6} \textnormal {-} 10 ^ {- 7} $. For this purpose, the minimum required number of events must be equal to $ N = 10 ^ {12} \textnormal {-} 10 ^ {14} $. According to \cite {Savin2014sva}, the Nuclotron is able to accelerate in one cycle up to $ 1.6 {\cdot} 10 ^ {11} $ polarized protons. We assume that on a very dense internal target, at least $ 10 ^ 9 $ of scattered particles will produce elementary flashes in scintillators surrounding the vacuum chamber in the target region. The complete statistics necessary for the experiment ( $ N = 10 ^ {12} \textnormal {-} 10 ^ {14} $ of recorded events will be collected for $ 10 ^ 3 \textnormal {-} 10 ^ 5 $ cycles of the NICA accelerator complex, which corresponds to quite reasonable working time (less than 1 year).

The dynamic distortion of the beam orbit, leading to the ~ interaction of particles with the ~ internal target, is supposed to be carried out in a fraction of a second, which is approximately $ 10 ^ 5 $ revolutions. Therefore, at each revolution about $ 10 ^ 4 $ events will be recorded, which will allow us to observe the azimuthal asymmetry of the  signals from the counters at the level of 1 \%. Since the azimuthal asymmetry is tens of percent at not very high energies and at 100\% transverse polarization of the beams, we can control the magnitude of the horizontal polarization of the particles during the interaction with the target with sufficient accuracy.

To provide acceptable parameters of the proton or deuteron beam circulating in ~ NICA, the beams must be cooled with the aid of the electron cooling device \cite {Parkhomchuk2018skc} created for this purpose. Estimated cooling time is 30 \ sec. The most critical parameter is the energy spread, especially for particles with ~ a large magnitude of the anomalous magnetic moment. For protons, it is desirable to have a spread of less than $ \sigma_\delta = 0.001 $, so that the effect of the high-frequency magnetic field of the flipper is more or less the same in strength for particles having different amplitudes of synchrotron oscillations. Indeed, the resonance harmonic value $ w $ of the  flipper, which is a short RF solenoid, is proportional to the value of the zero-order Bessel function $ J_0 (\xi) $, in which ~ the argument is $ \xi = \nu_0 a_\delta / \nu_s $, where $ \nu_0 $ is the spin frequency, $ a_\delta $ is the amplitude of synchrotron oscillations, and $ \nu_s $ is the frequency of synchrotron oscillations. For a coherent adiabatic rotation of spins in a ~ beam, it is necessary for modulation index $ \xi $ to be less than unity for all particles in the beam. For example, for protons with the ~ gamma factor $ \gamma = 3.3 $ we have: $ \nu_0 = 6 $ and $ \xi = 1 $ for $ \nu_s = 0.006 $ and $ a_\delta = 0.001 $. As follows from \cite {Parkhomchuk2018skc}, obtaining these parameters is a very realistic task. Note that the obtained restrictions on the energy spread of deuterons are  by an order of magnitude weaker due to the smallness of the absolute value of their anomalous magnetic moment, $ g = -0.143$. In this case, apparently, special preliminary cooling is not required.

\section{The main sources of systematics}

If the flipper axis does not coincide with the beam axis, which is a typical situation, then the field of flipper  can swing the orbit. This swing occurs synchronously with the precession frequency set by the flipper and leads to a dangerous systematics. This effect can be controlled by various methods. For example, you can study this background modulation on an unpolarized beam by measuring the amplitude and phase of the modulation of the count rate. A more radical way is to completely turn off the flipper after moving the spins to the horizontal plane and the beginning of the orbit shift to the target. In this case, the precession phase of free rotation is measured only by counters without reference to the flipper phase.
Though the proposed experiment can be carried out at the Nuclotron, the experiment in ~ one of the rings of the NICA collider has several advantages. First, in ~ Nuclotron there are no long free space for placing various recording equipment in them, while ~ in ~ NICA rings there are such very convenient spaces, for example, in the straight section for future SPD detector. Secondly, in ~ colliders the beam energy is much more stable since no acceleration is performed in them, ~ unlike the Nuclotron. Energy stability is very important, since knowledge of the resonant frequency of the spin precession is necessary at the level $ \Delta f / f = 10 ^ {- 4} \textnormal {-} 10 ^ {- 5} $ if we want to limit the harmonic value of the flipper to $ w \leqslant 10 ^ {- 4} $. For protons, it is also very important to provide  higher frequency of synchrotron oscillations in order to minimize the corresponding modulation indices for various particles in the beam.

\section{Parameters and technical implementation of the RF flipper}

\begin{figure}[htb]
\centering
\includegraphics[width=0.5\textwidth]{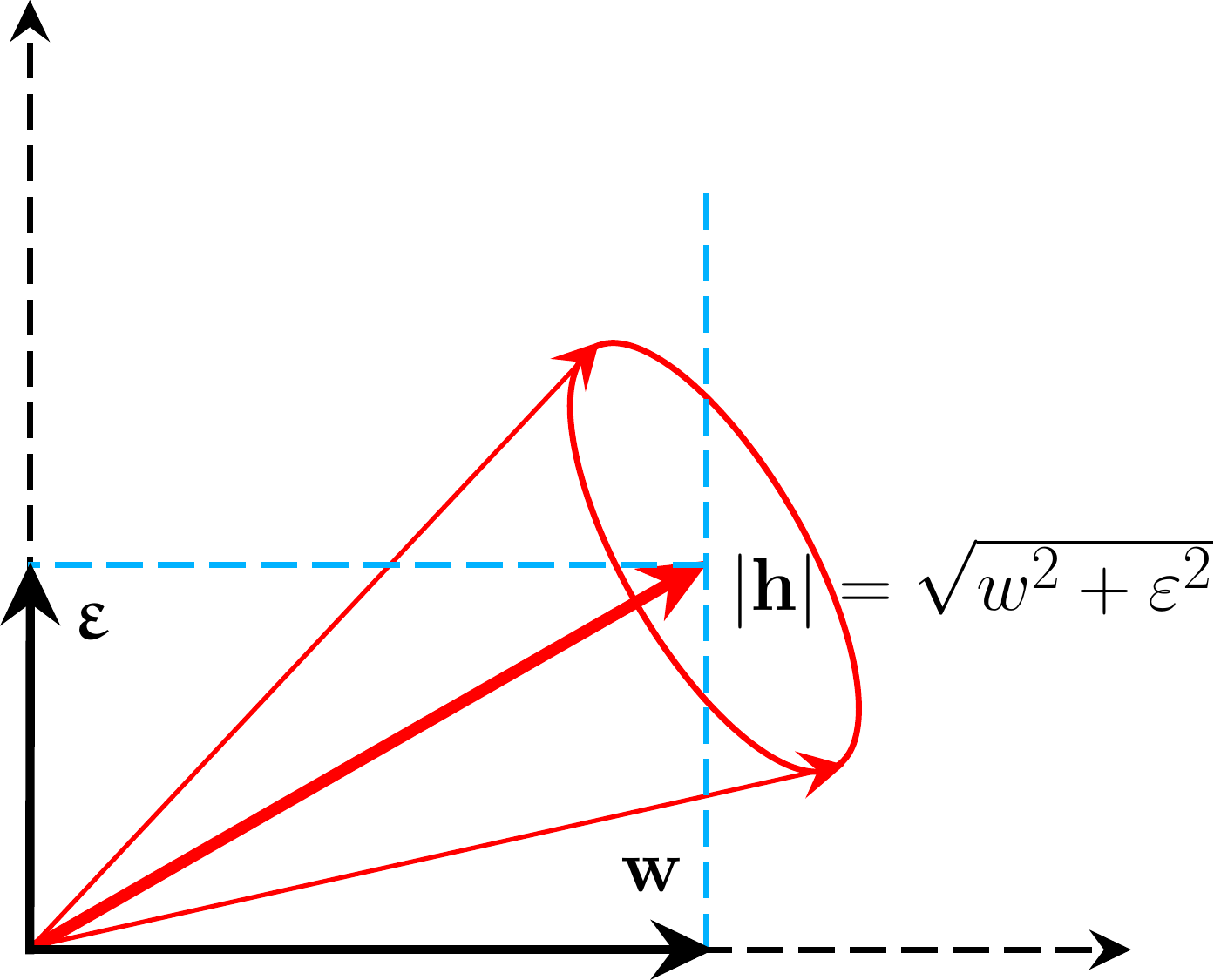}
\caption{Spin precession in a coordinate system rotating around a vertical axis (detuning vector $ \bm {\upvarepsilon} $ is directed along this axis) with ~ flipper frequency $ \nu $. In this frame of reference, which rotates synchronously with the vector $ \mathbf {w} $, the spin of an arbitrary particle rotates around the vector $ \mathbf {h} $, which is the sum of the vectors $ \mathbf {w} $ and $ \bm {\upvarepsilon} $; $ w = \lvert \mathbf {w} \rvert $ is equal to the amplitude generated by the circular harmonic of the flipper.}\label{fig:resframe}
\end{figure}

Fig. ~ \ref {fig:resframe} shows a vector diagram explaining the principle of adiabatic rotation into the ~ horizontal plane of the spins of particles having a vertical direction in the ~ initial state. The turn begins by turning on the flipper with ~ a large initial detuning $ \varepsilon = | \nu- \nu_0 | \gg w $ in the flipper frequency $ \nu $ relative to the resonant frequency $ \nu_0 $ and ends at a frequency close to ~ the exact resonance.

With an adiabatic change in the flipper frequency, the spin projection on the direction of the precession axis $ \mathbf {h} = \mathbf {w} + \bm {\upvarepsilon} $ is conserved, being an integral of motion. In this formula, the vector $ \mathbf {w} $ lies in the plane of the orbit and rotates around the vertical axis with a frequency of $ \nu $, and the vector $ \bm {\upvarepsilon} $ is directed perpendicular to the plane of the orbit.
Therefore, at the initial large detuning, when the precession axis almost exactly coincides with the ~ equilibrium vertical direction of all spins in the beam ~, the spin ensemble follows the precession axis. As a result, the spins are forced to come into a ~ horizontal precessing state by the time when the exact resonance sets in.
If the final detuning turns out to be nonzero, then the spins in the ~ laboratory frame of reference will rotate along a cone with the solution angle greater or less than $ 180^\circ $, i.e.,  with a nonzero vertical polarization component. Controlling its value with a polarimeter (by the left-right scattering asymmetry), we can very precisely tune the flipper to resonance with the free precession frequency. The resonance width is determined by the harmonic value of the flipper $ w $, because for $ \varepsilon = w $ the vector of the precession axis $ \mathbf {h} $ is obviously inclined at an angle of $ 45^ \circ $ to the ~ horizon, see ~ Fig.~\ref {fig:resframe}.

Depolarization of the beam due to the scattering of particles by the electrons of the residual gas and related energy diffusion cannot, according to our estimates, have a significant effect on the polarization lifetime in the scenario discussed here because of the fast process of the orbit shift toward the target. The depolarization time of the beam at the harmonic of the flipper $ w = 10^ {- 4} $, as our calculations show, is hours. Therefore, for deuterons it is sufficient  to provide the harmonic in the ~ range $ w = (3 \textnormal {-} 5) {\cdot} 10 ^ {- 5} $, and for protons at the level $ w = (1 \textnormal {-} 2) {\cdot} 10 ^ {- 4} $.

\begin{figure}[htb]
\centering
\includegraphics[width=0.5\textwidth]{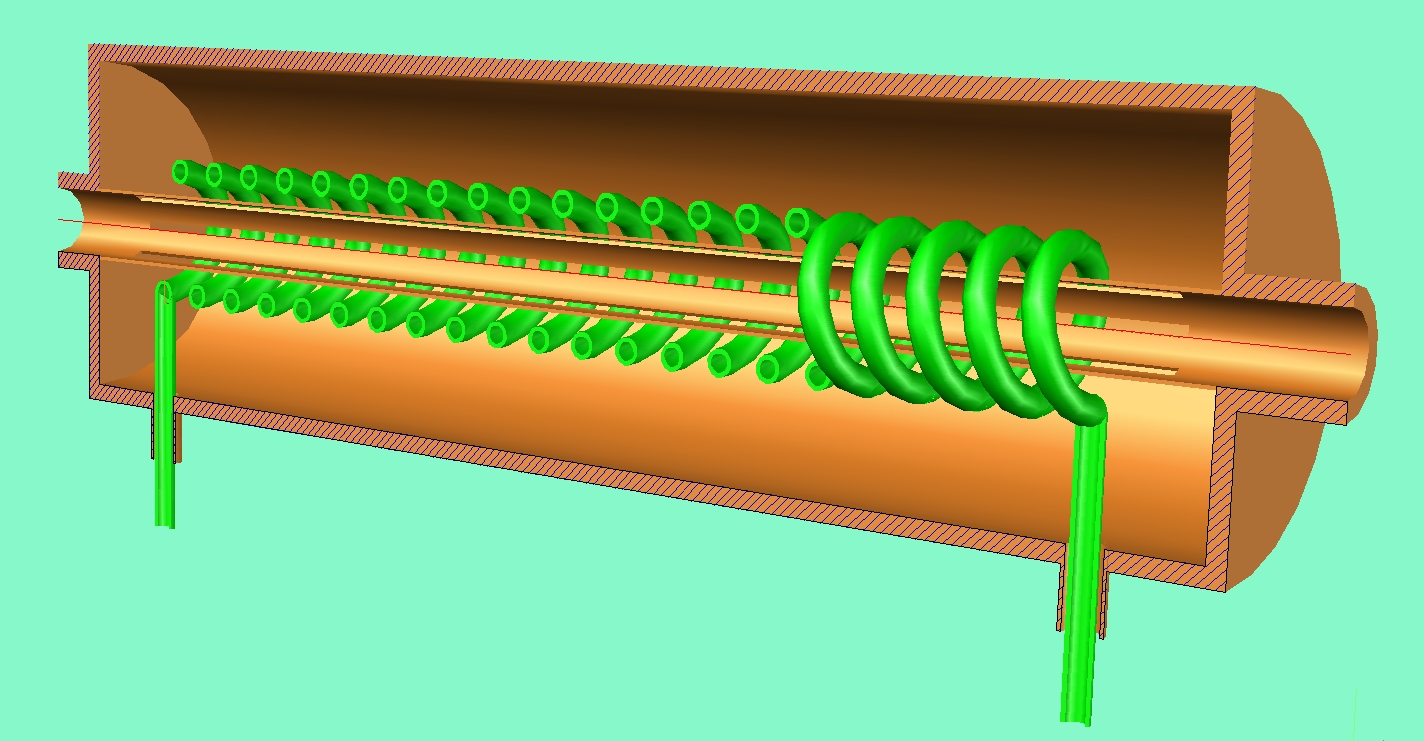}
\caption{The principal details of the high-frequency flipper device are: an RF solenoid coil, a cylindrical vacuum chamber and a copper liner with ~ several longitudinal slits-cuts, providing a low impedance of a complex volume filled with a spiral.}\label{fig:flipper}
\end{figure}

Let us discuss the technical problems of implementing an RF flipper. Fig.~\ref {fig:flipper} shows a sketch of a 3D model of a flipper. Its main element is a water-cooled copper spiral winding of an RF solenoid. Its ends outside the vacuum volume are grounded through capacitors of the desired rating, thus forming a series resonant circuit. Excitation of this circuit by the RF generator is performed as usual by a coupling loop. It is not shown in this figure.

Another important element of the flipper is the presence of a conductive copper liner with longitudinal slots located inside the solenoid. This screen provides a free propagation of the longitudinal currents of the beam image at all frequencies, as well as screening absence for the magnetic field of the solenoid on the axis of the system.

The table ~ \ref {tab:flipper-parameters} shows approximate parameters of the flipper when  operating at a frequency to be equal to half of the beam revolution frequency $ f_0 = 0.6 $ \, MHz.

\begin{table}[h]
\centering
\caption{The main flipper parameters for a deuteron momentum of 6.75 \,GeV/$ c $ assuming the value of the circular harmonic $ w = 10 ^ {- 4} $ ($ B\cdot\ell = 0.033 $ T$\cdot$ m)}\label{tab:flipper-parameters}
\begin{tabular}{@{}lcc@{}}
\toprule \multicolumn{1}{c}{Parameter} & Value & Units \\
\midrule Solenoid length & 2{,}5 & m \\
Coil diameter & 100 & mm \\
Liner aperture & 80 & mm \\
Chamber diameter& 300 & mm \\
Number of windings & 100 & \\
Coil inductance & 40 & mkH \\
Characteristic impedance& 75 & Ohm \\
Resistive impedance & 0.28 & Ohm \\
Quality factor & 270 & \\
Amplitude of the magnetic field & 0.013 & T \\
Coil current & 262 & A \\
Inductive voltage  & 1.5 & kV \\
Active losses power  & 10 & kW \\
\bottomrule
\end{tabular}
\end{table}

\section{Method of the total cross section measuring}

In this method, the total cross sections for the interaction of longitudinally polarized proton or deuteron beams with an unpolarized external target are measured for opposite signs of particle helicities in the beams. This approach was used in the experiment \cite {Lockyer1984pg}, in which the beam intensity was measured before and after the target, which was 81 cm of water. In the experiment \cite{Lockyer1984pg}, the stationary polarization of the proton beam in the ~ ring was vertical. It was converted into ~ longitudinal by rotating the velocity vector by the required angle in the  output channel. Therefore, the experiment was conducted at a fixed energy of 5.1\, GeV,  which corresponds to spin rotation  to $ 90 ^ \circ $.

We now discuss the advantages and disadvantages of setting up an experiment with a ~ beam being extracted  from a NICA storage ring into a ~ special channel, where it interacts with a ~ dense target, and the recording equipment measures the beam transmission coefficient through the target. Firstly, there is a great deal of freedom in the choice of the location geometry and type of a target. The main advantage of such a more traditional formulation is that the total cross section is measured here, and not some part of it, as in the ~ version with an internal target. The transmission coefficient is measured by two integrated counters, based, for example, on the registration of the secondary emission of electrons from the material of dozens of membranes interlayered with collecting electrodes. Such a counter is in our development stage.

Unlike the experimental setup used in ~ \cite {Lockyer1984pg}, we propose, as in our previous approach, to use a flipper to control the direction of the polarization vector in the NICA ring before extracting it to the ~ channel. There are several considerations for this. Firstly, turning the deuteron spin from vertical to ~ horizontal with the help of rotation of the velocity vector, as was done in ~ \cite {Lockyer1984pg} with ~ protons, is almost impossible due to the small magnetic moment of the deuterons. Secondly, it seems to us that it is necessary to preserve complete freedom in ~ controlling the orientation of the spin at the ~ location of the target. This is important from the point of view of combating systematics when determining the dependence of the coefficient of passage of a particle beam through a target on the longitudinal spin component. The point is that, as explained above, the presence of even a small transverse component of the spin can lead to a ~ change in this coefficient due to errors in the counter alignment relative to the beam axis. It seems to us that the only way to see this dependence is to directly measure the modulation value of the beam passing through the target with opposite signs of the transversely oriented spins of the beam. Flipper will allow this to be done in ~ special control extractions of beam bunches with ~ any desired spin orientation, in contrast to the static approach used in all previous experiments.

Note that using of a  transversely polarized beam are also very important for suppressing the systematics, because it allows one to take into account the asymmetry, not related to ~ longitudinal polarization (with weak interaction), but related to the inaccuracy of the equipment alignment. This information can be taken into account at offline processing of the measurements. 

Unfortunately, the experimental scheme under discussion, in addition to the advantages, has some disadvantages. First, the transmission coefficients with ~ different signs of the longitudinal component of the spins are compared using the data obtained in ~ different extractions alternating in the sign of helicity, and not inside one, as in a ~ scheme with precessing polarization and an internal target. In fact, with good repeatability of the parameters of the beam emitted into the ~ channel, this effect should not significantly increase the fluctuation of the results.
Secondly, in the ~ design of the NICA collider, the special beam line for extracted particles is not provided, although its creation, apparently, is not too difficult task.

\section*{Conclusion}
We have discussed two approaches to  investigate the effect of $ P $-parity violation in the interaction of polarized proton or deuteron beams with an ~ unpolarized target.
In one of these approaches, it is suggested to study the scattering of a polarized beam of protons or deuterons on a thick internal target. In this formulation, the spin of a circulating particle beam in one of the NICA rings translates into a mode precessing in the horizontal plane using an RF flipper. The effect of parity violation will be studied by measuring the correlation of the cross section for the interaction of particles and the direction of their spins.

In an alternative approach, the flipper translates the spins of particles into a ~ horizontal plane and the beam is extracted  into a ~ channel in a ~ certain precession phase. In this more traditional experimental setup, the total cross section of  particles  passing  through a dense target is measured depending on the sign of the polarization helicity of the beam.

Both approaches  have their pros and cons, and a final choice between them remains to be made  during a detailed simulation of the experiment.

Backgrounds also require special consideration. For example, it is important to study the background arising from the contribution to the ~ measured cross section of decays of longitudinally polarized hyperons generated on the target. This decays can significantly affect the size of the measured parity violation effect \cite {Lockyer1984pg}.

\subsection*{Acknowledgments}
We are grateful to Yu.M. Zharinov and A.V. Otboev for technical assistance.
This work is supported  by RFBR Grant No. 18-02-40092 MEGA.

%\bibliographystyle{pepan}
%\bibliography{Koop-NICA-flipper}

\end{document}